\documentclass[conference,12pt,onecolumn,draftclsnofoot,]{IEEEtran}
\IEEEoverridecommandlockouts
\usepackage{cite}
\usepackage{amsmath,amssymb,amsfonts}
\usepackage{algorithmic}
\usepackage{graphicx}
\usepackage{textcomp}
\usepackage{xcolor}
\def\BibTeX{{\rm B\kern-.05em{\sc i\kern-.025em b}\kern-.08em
    T\kern-.1667em\lower.7ex\hbox{E}\kern-.125emX}}

\usepackage{amsfonts}
\usepackage{algorithmic}
\usepackage{caption}
\usepackage{subcaption}

\usepackage[ruled,vlined,linesnumbered]{algorithm2e}

\begin{document}

\title{EmbAu: A Novel Technique to Embed Audio Data using \\
Shuffled Frog Leaping Algorithm

\thanks{Corresponding author: Ankit Chaudhary (dr.ankit@ieee.org)}}

\author{\IEEEauthorblockN{ Sahil Nokhwal }
\IEEEauthorblockA{\textit{Dept. Computer Science} \\
\textit{University of Memphis}\\
Memphis, TN, USA \\
snokhwal@memphis.edu}
\and

\IEEEauthorblockN{ Saurabh Pahune }
\IEEEauthorblockA{\textit{Automation analyst} \\
\textit{Cardinal Health}\\
Dublin, Ohio, USA \\
saurabh.pahune@cardinalhealth.com}
\and

\IEEEauthorblockN{ Ankit Chaudhary}
\IEEEauthorblockA{\textit{Dept. Computer Science} \\
\textit{University of Missouri}\\
St Louis, MO, USA \\
dr.ankit@ieee.org}
}

\thispagestyle{plain}
\pagestyle{plain}
\maketitle
\begin{abstract}
 The aim of steganographic algorithms is to identify the appropriate pixel positions in the host or cover image, where bits of sensitive information can be concealed for data encryption. Work is being done to improve the capacity to integrate sensitive information and to maintain the visual appearance of the steganographic image. Consequently, steganography is a challenging research area.

In our currently proposed image steganographic technique, we used the Shuffled Frog Leaping Algorithm (SFLA) to determine the order of pixels by which sensitive information can be placed in the cover image. To achieve greater embedding capacity, pixels from the spatial domain of the cover image are carefully chosen and used for placing the sensitive data. Bolstered via image steganography, the final image after embedding is resistant to steganalytic attacks. The SFLA algorithm serves in the optimal pixels selection of any colored (RGB) cover image for secret bit embedding. Using the fitness function, the SFLA benefits by reaching a minimum cost value in an acceptable amount of time. The pixels for embedding are meticulously chosen to minimize the host image's distortion upon embedding. Moreover, an effort has been taken to make the detection of embedded data in the steganographic image a formidable challenge.	

Due to the enormous need for audio data encryption in the current world, we feel that our suggested method has significant potential in real-world applications. In this paper, we propose and compare our strategy to existing steganographic methods.
\end{abstract}

\begin{IEEEkeywords}
Steganography, Cipher, Data hiding, Data Security, Confidentiality, Privacy, Cryptography,  SFLA algorithm
\end{IEEEkeywords}

\maketitle

\section{Introduction}\label{sec:intro}
Increases in both home and office computer use have led to a meteoric rise in the number of people connected to the internet over the past two decades, resulting in an increase in privacy concerns, security problems worldwide\cite{tang2015image, mohsin2018real, naji2009novel, zaidan2009securing, zaidan2010differences}, cybercrime, as criminals attempt to steal high-value assets, which is now in the form of data. In recent years, there have been a rising number of eavesdropper attacks and information thefts. Cybersecurity experts and researchers on the other hand, aim to protect network data transmission from unauthorized interception by inventing the appropriate method, many of which are based on cipher algorithms, where data is first encrypted by the sender, then transmitted and must be decrypted by the receiver in order to get the confidential message. In this way, access to the data is restricted to only those who know how to decrypt it. Despite the fact that data encryption has yielded outstanding results in the past and continues to do so, there is a significant research interest in enhancing the present encryption methods.

Cipher text is inconspicuous, and thus attracts the attention of unauthorized users, preventing even authorized recipients from receiving it. A more nuanced approach would be to disguise the information in a manner that its presence cannot be detected by unauthorized users. These methods of information hiding are relatively new areas of research interest and yet have shown considerable promise. A method of hiding information involves hiding data from one file into another, the files may or may not have the same format, for instance, hiding an image in another image or audio data in an image.

Steganography as a method of data hiding boasts an incredible storage capacity \cite{majeed2009novel, zaidan2010novel, eltahir2009high, al2011securing, othman2009extensive, al2010overview}. When an image is chosen as the medium of information hiding, the image into which information is embedded is called the host image or cover image. After data embedding, the final image is called a steganographic image \cite{zaidan2009approved, raja2013efficient, naji2009stego, ahmed2010novel, islam2010new}. Steganography is performed for three purposes: 1) The first is to hold or deliver payload (data) \cite{al2010overview, raja2013efficient, naji2009stego, ahmed2010novel, islam2010new, zaidan2009new, jalab2009frame, abomhara2011experiment, jalab2010new, mahmoud2010optimization, yahya2010new}, and the second is to protect this payload from being accessed by people other than the intended users, and 3) is to secure the overall system \cite{cheddad2010digital, hmood2010capacity, majeed2009novel, abomhara2010suitability}. Thus, there is much research interest in methods of embedding data into images, while optimizing for higher payloads \cite{uma2017performance, naji2009novel, zaidan2010stegomos} without compromising much on image quality. To ascertain that a desired standard of quality is maintained, we measure image visual quality using peak signal-to-noise ratio (PSNR) and Structural Similarity Index (SSIM) metrics \cite{hameed2009novel, hameed2010accurate}. Also, the storage capacity of an image is measured in bits-per-pixel. 
In this research, we present a novel approach for data embedding in images, in which we seek to determine the ideal sequence of pixels in which to embed data while maintaining image quality.
Here we build upon the technique proposed in \cite{gupta2018metaheuristic} and demonstrate improved results. Our suggested method gives us control over the convergence speed of data embedding and thus can be faster. This proposed method is of special interest also due to it being based on swarm intelligence and thus resulting in optimal pixels selection for data embedding.
The remaining sections in this paper are laid out as follows: section \ref{sec:related_work} explains the related work, 
followed by explanation of the SFLA algorithm in section \ref{sec:sfla_algo}, section \ref{sec:proposed_tech} provides details about the proposed technique, and section \ref{sec:experimental_results} shows the experimental results. In section \ref{sec:conclusion}, the conclusion of our research has been discussed.

\begin{figure*}[htbp]
    \centering
    \includegraphics[width=0.8\linewidth]{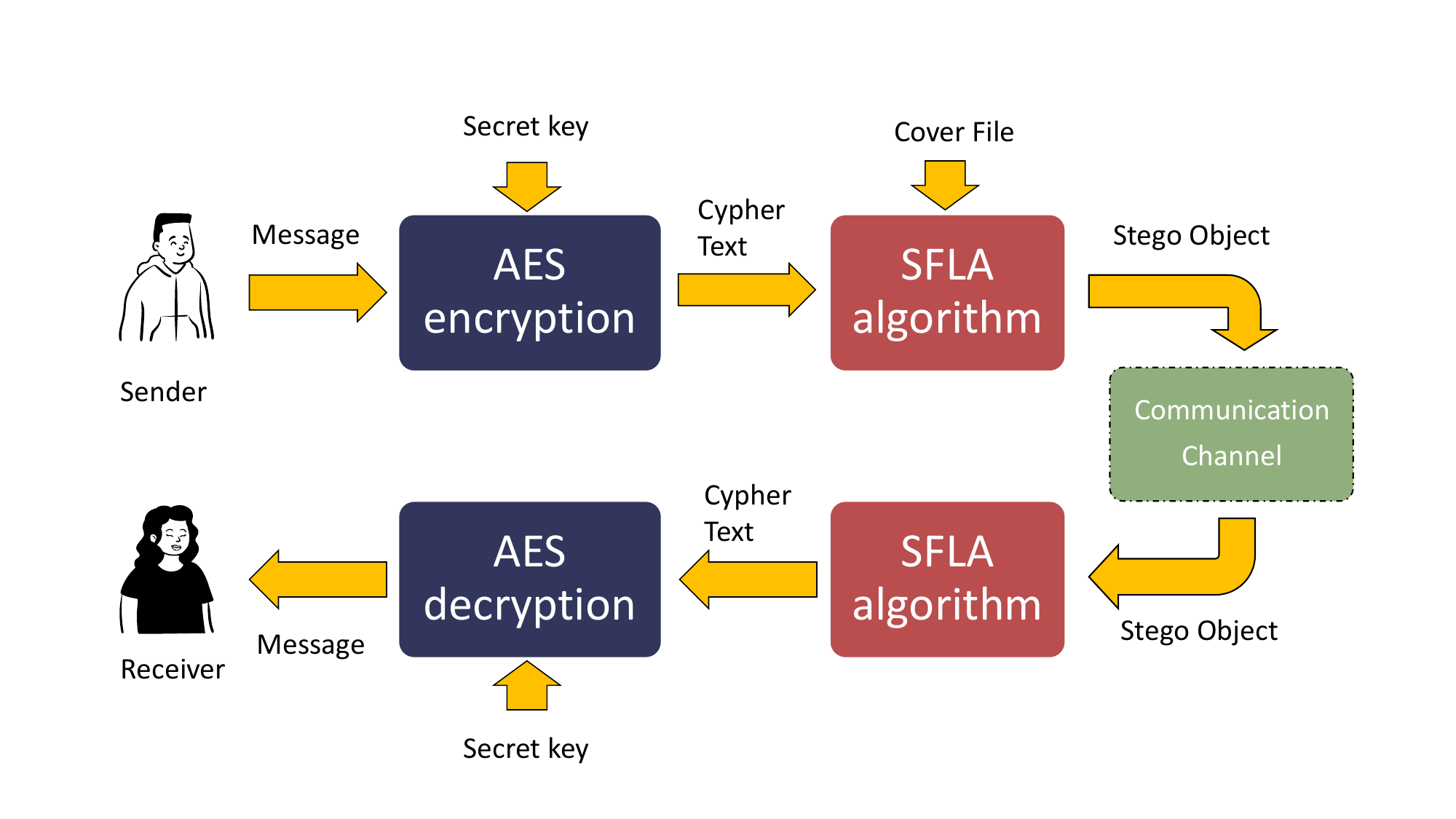}
    \caption{Architecture of proposed steganographic method}
\end{figure*}

\section{Motivation}\label{sec:motivation}
Despite the potential importance of image steganography, it has been largely overlooked in favor of more well-known security methods like encryption and watermarking. Limited  robustness, stealth, and payload capacity of current steganographic techniques are three areas where the current literature is lacking that are being investigated here. Lack of security, a small payload, poor resilience, and significant computing costs are the salient drawbacks of existing approaches. 

\begin{algorithm}[!htb]
    \DontPrintSemicolon
    \SetKwFunction{FSfla}{SFLAalgorithm}
    \SetKwProg{Fn}{Function}{:}{}
    \Fn{\FSfla{$\mathcal{C}, \mathcal{A}$}}{
        \KwIn{$\mathcal{C}$ - Cover image,
        $\mathcal{A}$ - audio file (optional)
        }
        \KwOut{$\mathcal{Q}$ sequence of optimal pixel locations}
        \BlankLine
			\textbf{Begin:}
			$\mathcal{F} \gets$ Generate virtual population of frogs \;
			\While{Termination condition not met}{
				Arrange the $ \mathcal{F} $ frogs in decreasing order of their fitness. \\
				Let $ X_{g} $ denote the location of the frog with the best global performance. \\
				Form $ m $ memeplexes, each containing $ n $ frogs. \\
				Allocate frog $ i $ to memeplex $ r = i \mod m $ \\
				
				\For{each memeplex}{
					Create a submemeplex of $ q $ chosen at random frogs \;
					\For{each submemeplex}{
						The submemeplex's best and worst frogs' positions will be denoted by $ X_{b} $ and $ X_{w} $, respectively. \;
						Determine worst frog's new location ($ X_{new1} $) using
						$ X_{new1} = X_{w} + r(X_{g} -X_{w}) $; where $ r $ is an arbitrary integer ranging from 0 to 1 generated uniformly \;
						
						\If{Performance of $ X_{new1} $ is inferior than $ X_{w} $}{
						    Determine the wrost frog's new location ($ X_{new2} $) using \;
							$ X_{new2} = X_{w} + r(X_{g} -X_{w}) $; 
							
							\If{Performance of $ X_{new2} $ is inferior than $ X_{w} $}{
    							Create a replacement frog  $ X_{w} $ at random.
    							}
							\Else{
								Substitute $ X_{w} $ for $ X_{new2} $
							}
							}
						\Else{
							Substitute $ X_{w} $ for $ X_{new1} $
						}
						Arrange memeplex's $n$ frogs in decreasing order. \;
					    Iterate for $N$ iterations
				}
				}
				The memeplexes would be thoroughly merged
			}
			$\mathcal{Q} \gets$ return the optimal sequence of pixels
			}
		\caption{SFLA algorithm to find a sequence of optimal pixels}
		\label{algo:sfla}
	\end{algorithm}

\section{Related work}\label{sec:related_work}
Most existing literature on steganographic techniques is based on the Particle Swarm Optimization (PSO) algorithm\cite{raja2013efficient}. These works outline PSO-based techniques that optimally find pixels in which secret information is to be  embedded while retaining image quality. Authors in  \cite{nickfarjam2014image, bedi2013using} discuss a well-performing technique that involves embedding the most significant bit of secret data in the least significant bit of an image. In \cite{al2017new, el2015new}, use adaptive neural networks to cut down on noise in the steganographic image. Authors in \cite{rajeswari2017application} proposed a method that utilizes wavelet decomposition. In \cite{nipanikar2018sparse}, a technique is detailed that hides data in images using wavelet transforms, and  uses the PSO algorithm to find optimal pixels for such hiding. In \cite{sajasi2015adaptive}, the proposed approach is a hybrid of optimal chaotic-based encryption method and noise visibility function which boasts of high payloads. \cite{li2018steganography} makes use of the difference between neighboring pixels of the host image, and uses the modulo function for data hiding. Careful modification of digital image pixels and hiding of sensitive information in the least significant bits (LSBs) of image components, the authors of \cite{khamrui2013genetic, lee2010novel}, suggested a variety of effective data concealing techniques. Authors of \cite{khan2023asymmetric} had previously explored the random LSBs-based hiding, later built a steganographic approach with variable least significant bits (VLSB). Using several methods detailed in \cite{hong2009reversible, hsu2010probability, khan2023asymmetric} successfully implemented variable-LSB (VLSB) steganography. For the data concealing and retrieval processes, they devised a prediction system and decision structure for selecting the insertable grouping of data bits. Medical raw data encoding  \cite{sarma2023power}, image compression \cite{banafshehvaragh2023intrusion}, and internet of things security \cite{li2023simple}, are also common applications of absolute moment block truncation coding (AMBTC) based approaches. Confidential data security was further enhanced by using block-based data-hiding methods, and novel data-hiding strategies are presented to assure data safety. Some of the promising block-based methods were proposed in \cite{khan2011implementation, irfan2014analysis}.


\begin{algorithm}[!htb]
    \DontPrintSemicolon
    \SetKwFunction{FSfla}{EmbeddingData}
    \SetKwProg{Fn}{Function}{:}{}
    \Fn{\FSfla{$\mathcal{C}, \mathcal{K}, \mathcal{A}$}}{
        \KwIn{$\mathcal{C}$ - Cover image to hide the audio data,
        $\mathcal{K}$ - Secret key used for encrypting audio data, 
        $\mathcal{A}$ - audio file
        }
        \KwOut{$\mathcal{S}$ steganographic image}

        \BlankLine
        
			$\mathcal{B} \gets$ Read audio data and convert it into bits \;
			$\mathcal{B} \gets$ Encrypt audio bits using AES algorithm \;
			$\mathcal{F} \gets$ Generate virtual population of frogs \;
			$\mathcal{Q} \gets$ gets optimal sequence of pixels to hide data using the SFLA algorithm~\ref{algo:sfla} \;
			$\mathcal{S} \gets$ Embed the encrypted audio bits into the cover image and return steganographic image
			}
		\caption{Proposed algorithm for embedding data}\label{algo:hidingData}
	\end{algorithm}

\begin{algorithm}[!htb]
    \DontPrintSemicolon
    \SetKwFunction{FSfla}{RetrievingData}
    \SetKwProg{Fn}{Function}{:}{}
    \Fn{\FSfla{$\mathcal{S} , \mathcal{K}$}}{
        \KwIn{$\mathcal{S}$ - Steganographic image to retrieve the audio data,
        $\mathcal{K}$ - Secret key used for decrypting the audio data
        }
        \KwOut{$\mathcal{A}$ audio file}
        \BlankLine
			$\mathcal{F} \gets$ Generate virtual population of frogs \;
    		
    		$\mathcal{Q} \gets$ gets optimal sequence of pixels to hide data using the SFLA algorithm~\ref{algo:sfla} \;
    		
			$\mathcal{A} \gets$ Retrieve encrypted audio data \;
			$\mathcal{A} \gets$ Decrypt audio bits using AES algorithm \;
			$\mathcal{A} \gets$ Make a audio file and return  \;
			}
		\caption{Proposed algorithm for retrieving data}\label{algo:retrievingData}
	\end{algorithm}

\section{The SFLA algorithm}
\label{sec:sfla_algo}
Based on the flocking behavior of frogs, a swarm-based, memetic algorithm SFLA was designed that performs a local and global search to find an optimal or near optimal solution. The SFLA algorithm makes use of both stochastic and deterministic techniques. Because of its deterministic approach, the SFLA algorithm is able to make good use of the information provided by the local search values to direct the heuristic exploration, much as is done in the PSO algorithm. The search process is adaptable and reliable because of the random variables. Initially, a considerable number of random frogs $\mathcal{F}$ are released into the marsh to swarm everywhere. The population is split up into a few separate parallel groups, known as memeplexes, and each of these groups is allowed to grow independently so that they may explore the space in a wide range of ways. The frogs undergo a memetic evolution as a result of being influenced by the concepts of other frogs while they are participating in each memeplex. The effectiveness of every meme is improved via the process of memetic evolution, which also increases a frog's overall effectiveness toward achieving a desired result. Frogs that have superior memes (ideas) are allowed to contribute greater amounts to the production of new ideas than frogs that have poor ideas. This is done to guarantee that the influenced procedure remains competitive. A competitive edge may be gained by superior ideas via the use of a triangle probability distribution in the selection of frogs. Frogs may update their memes throughout the evolution by taking the most effective ideas from the whole population or just the memeplex. Updates across the memotype(s) that occur incrementally correlate to the size of a leaping stride, and the appearance of a new meme represents the frog's new location. Each frog is given the opportunity to improve itself before being released back into the memeplex. There is no delay in using the acquired knowledge to again improve the condition. This method differs from a genetic algorithm in that it does not need the whole population to be altered prior to new findings. Instead, this method provides quick access to newly discovered information. Here, the graphic depiction of frogs is utilized as a comparison, but in terms of the dissemination of ideas, the parallel may be used to groups of scientists or scholars enhancing an idea or experts incrementally improving a model. When the meta-heuristic evolutionary cycles reach a specific threshold, the memeplexes are compelled to interbreed, and novel memeplexes get generated as a result of a shuffling procedure. After coming into contact with frogs from various parts of the such swamp, the memes benefit from this mixing and matching. The process of looking for a potential solution is accelerated by the shuffling of frogs, which also assures that there is no local prejudice in the progress of any special relevance.

\section{Proposed Technique}\label{sec:proposed_tech} 
The proposed technique has been divided into two sections, i.e., embedding audio data into the cover image; and retrieving the already hidden data from the steganographic image. 

\subsection{Embedding audio data using the SFLA algorithm}\label{sec:embedding}
Our proposed technique aims to locate the sequence of the optimal pixels for embedding audio data so that distortion in the cover image is minimized. The steps for our proposed technique are
\begin{enumerate}
    \item First read the confidential audio file that needs to be stored in the cover image in its decimal format
    \item Encrypt the decimal values obtained from the audio file using the AES encryption algorithm which outputs 128-bit data
    \item In order to save this encrypted data in the cover image, these encrypted 128-bits data would be transformed into a series of binary bits
    \item The cover image's pixels are all transformed to their 8-bit equivalents
    \item A custom objective function is developed based on the multi-objective scenario. It is a metric used to measure the quality of a steganographic image compared to the original signal. PSNR is calculated as the ratio of the maximum possible power of a signal to the power of the noise that affects the fidelity of its representation. The PSNR value is expressed in decibels (dB) and it provides a quantitative measure of the amount of distortion present in the steganographic signal. The higher the PSNR value, the better the quality of the steganographic image. 
    
    The Structural Similarity Index (SSIM) is a metric used to measure the similarity between two images. It is a structural similarity measure that takes into account the structural information and luminance of the images being compared. It attempts to model the human visual system by measuring the changes in structural information, luminance, and contrast between the two images. SSIM is calculated by comparing the luminance, contrast, and structural information of two images. It outputs a value between -1 and 1, where a value of 1 indicates that the two images are identical, and a value of -1 indicates that the two images are completely dissimilar. 
    
    Mean Squared Error (MSE) is a commonly used metric to measure the average squared difference between the estimated values and the true values of two images. It measures the quality of a steganographic image. The calculation is done by taking the average of the squared differences between the corresponding pixels or samples in the cover and steganographic images. 
    
    \begin{equation*}
        MSE = \sum_{i=0}^{m-1}\sum_{j=0}^{n-1}\left[ \mathcal{X}(i, j) - \mathcal{Y}(i, j)\right]^2
    \end{equation*}
    
    where $(i, j)$ represents the $i^{th}$ and $j^{th}$ pixel of the cover $\mathcal{X}(i, j)$ and steganographic $\mathcal{Y}(i, j)$ image, and $|m| = |n|$ as the total pixels of the cover and steganographic image are equal.
    \\
    
    The equation for PSNR can be formulated as
    \begin{equation*}
        PSNR = 10\times \log \left(\dfrac{MAX^2_C}{MSE}\right) 
    \end{equation*}
    
    \begin{equation*}
        PSNR = 10 \times \log\left(MAX^2_C\right) - 10 \times \log\left(MSE\right)
    \end{equation*}  
    
    Where $MAX_C$ represents the maximum pixel value of the cover image. 
    Therefore, a general fitness function can be formulated as
    
    \begin{equation*}
        Z(\mathcal{X}, \mathcal{Y})     = \alpha \times SSIM(\mathcal{X}, \mathcal{Y}) + (1 - \alpha) \times \dfrac{PSNR(\mathcal{X}, \mathcal{Y})}{100}
    \end{equation*}
    Here, $\mathcal{X}$ and $\mathcal{Y}$ represent two images that are being compared. In our work, $\mathcal{X}$ represents a steganographic image, and $\mathcal{Y}$ represents a cover image. As the cover image $\mathcal{Y}$ is fixed during the entire process of embedding, the fitness function can be formulated as
    \begin{equation}\label{eq:objFun}
        Z(\mathcal{S}, \mathcal{C})     = \alpha \times SSIM(\mathcal{S}, \mathcal{C}) + (1 - \alpha) \times \dfrac{PSNR(\mathcal{S}, \mathcal{C})}{100}
    \end{equation}
   Here the value of $\alpha$ is taken as 0.5.

    Here the value of SSIM gives the universal image quality index, which is described as follows
    \begin{equation*}
        Q = \dfrac{4 \sigma_{uv} \hat{u}\hat{v}}{ (\hat{u}^2 + \hat{v}^2) (\sigma_{u}^2 + \sigma_{v}^2) }
    \end{equation*}
    
    Here, the steganographic image's window size is represented by $u$ and $v$, and its average is represented by $\hat{u}$ and $\hat{v}$.
    The variance of $u$ and $v$ is represented by $\sigma^u_2$ and $\sigma^v_2$, and their covariance is represented by $\sigma_{uv}$.
    
    The metric takes Human Visual System (HVS) into account. The HVS's model distortions comprise of three different elements: luminance distortion, loss of correction, and contrast distortion. They are mathematically modeled as follows: 
    
    \begin{equation*}
       Q = \dfrac{2 \sigma_{uv}}{\sigma_{u} \sigma_{v}} \times  \dfrac{2\hat{u}\hat{v}}{\hat{u}^2 + \hat{v}^2} \times  \dfrac{\sigma_{u}\sigma_{v}}{(\sigma_{u}^2 + \sigma_{v}^2)}
    \end{equation*}

    The first element exhibits some linear correlation and has an inclusive dynamic range of -1 to 1. The dynamic range of the second element is between 0 and 1 inclusive, and it displays the brightness $u$ and $v$. The range of the third element, which represents how comparable the images' contrasts are, is between 0 and 1.
    Consequently, the dynamic range of $Q$ is between -1 and 1. When the denominator turn in the QUI approaches zero, then results become unstable. To avoid this problem, we use a generalized metric called SSIM and it tells how similar two images are using the following equation
    \begin{equation*}
        SSIM (u,v) =  \dfrac{(2\hat{u}\hat{v} + c_1)(2\sigma_{uv} + c_2)}{(\hat{u}^2 + \hat{v}^2 + 1) (\sigma_{u}^2 + \sigma_{v}^2 + c_2)}
    \end{equation*}
    
    Where $u$ and $v$ indicate the window size for the original image and steganographic image, respectively, and $u$ and $v$ denote the mean of $u$ and $v$. $\sigma^u_2$ and $\sigma^v_2$ denote the variances of $u$ and $v$, whereas $\sigma_uv$ indicates their covariance.

    \item \emph{Population initialization:} Initial population generation of frogs is based on random initialization and then dividing the population evenly by the required number of memeplexes. Generate random $\mathcal{F}$ frogs $\mathcal{U}(1)$, $\mathcal{U}(2)$, $\mathcal{U}(3)$,$\mathellipsis$, $\mathcal{U}(F)$  those lies in solution space $\Omega \in \mathbb{R}^{d}$, where $\mathcal{D}$ represents the total number of determinants. Therefore, a vector of values $\mathcal{U}(i)$ = $\mathcal{U}(2)$, $\mathcal{U}(3)$,$\mathellipsis$, $\mathcal{U}(F)$ is described for the $i^{th}$, decision variable stands in for the $i^{th}$ frog. Determine the $\mathcal{f}(i)$ value to every frog in the set $\mathcal{U}(i)$.

    \item \emph{Initialize memeplexes:} The number of frogs $\mathcal{F}$, is divided into memeplexes, i.e., $m$ and each memeplex has $\eta$ frogs, where $\eta$ is an integer. Hence, the equation depicting the aggregate number of samples $\mathcal{F}$ in the swamp can be formulated as
    \begin{equation*}
      \mathcal{F} = m \times \eta
    \end{equation*}
      
    Here $\mathcal{F}$ depicts the total number of frogs in the initial population, $m$ represents the total number of memeplexes, and $\eta$ corresponds to the number of frogs within every memeplex.
     
    \item \emph{Rank frogs:} Each frog's fitness is determined and used to arrange the population in memeplexes in descending order and in a round-robin manner.
    \item \emph{Perform Search:} A local search is performed in each memeplex based on a memetic algorithm. After a few iterations, a few frogs jump from one memeplex to another, then the fitness of each frog is recalculated, after which they rejoin  the memeplex based on their fitness.
    Thus, a new population is constructed. These steps are repeated until terminating conditions are met or an optimal solution is found.
    
    \item Using the SFLA approach, determine the best set of pixels to mask the data in the host image.
    \item Utilize the 1-LSB approach to embed the encrypted data bits in the cover image with minimal image distortion
\end{enumerate}

 \subsection{Retrieving the embedded audio data from steganographic image}
    \begin{enumerate}
    \item Resize the steganographic color image of 3 channels (m x n x 3) into m' x n' image (2-dimensional)
    \item A custom objective function based on SSIM and PSNR values is developed as per equation~\ref{eq:objFun}
    \item \emph{Population and memeplexes initialization:} The initial population generation of frogs is done as per as the section~\ref{sec:embedding}
    \item After configuring all settings, the SFLA algorithm is executed to minimize the objective function, yielding an optimal sequence of pixels for retrieving audio data
    \item Retrieve the encrypted data based on the sequence obtained from the previous step
    \item Decrypt the data into its binary representation using the AES algorithm
    \item Make an audio file using the decrypted binary bits
\end{enumerate}

\begin{figure}[!ht]
    \centering

    \begin{minipage}[b]{0.5\textwidth}
        \centering
        \includegraphics[width=0.85\textwidth]{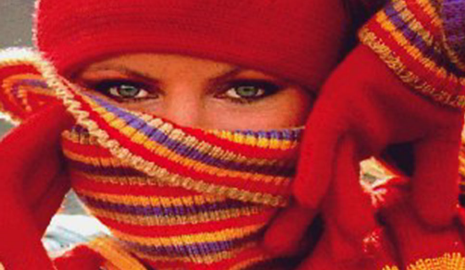}
        \caption{Before}
    \end{minipage}
    \begin{minipage}[b]{0.5\textwidth}
        \centering
        \includegraphics[width=0.85\textwidth]{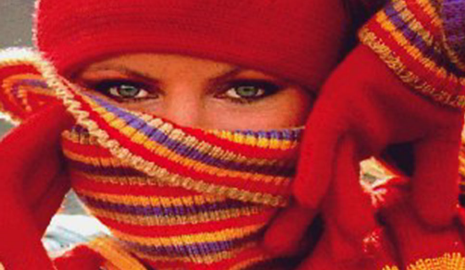}
        \caption{After}
    \end{minipage}

    \caption{Before and after steganography - Fashion}
\end{figure}

\begin{figure}[!ht]
    \centering
   \begin{minipage}[b]{0.5\textwidth}
        \centering
        \includegraphics[width=0.85\textwidth]{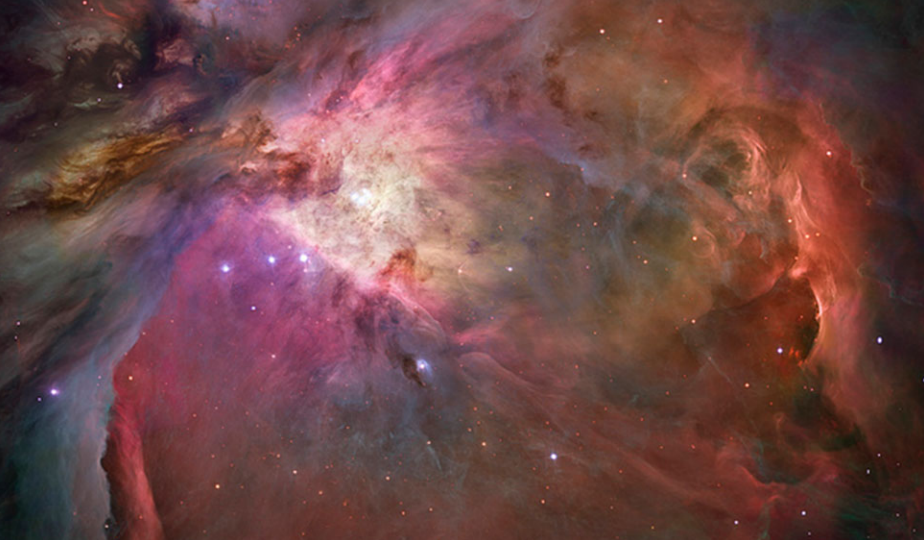}
        \caption{Before}
    \end{minipage}
    \begin{minipage}[b]{0.5\textwidth}
        \centering
        \includegraphics[width=0.85\textwidth]{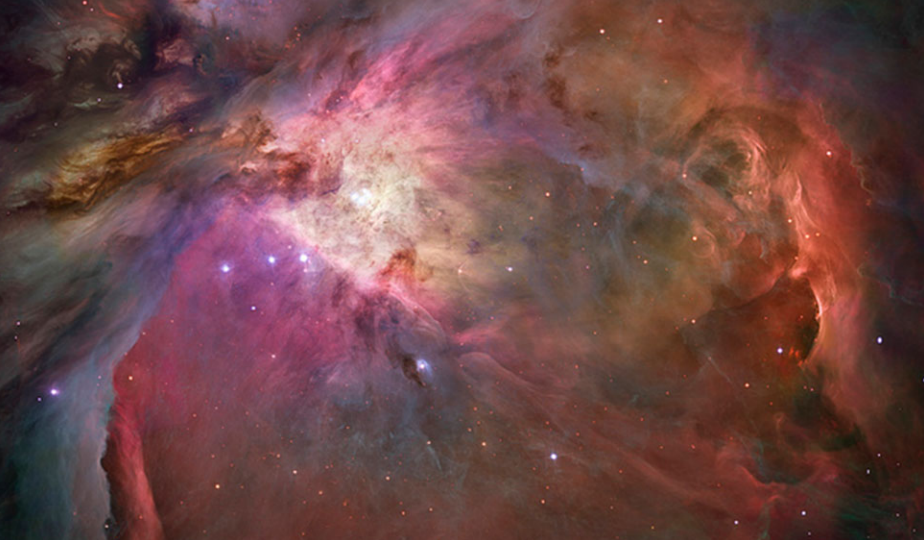}
        \caption{After}
    \end{minipage}
     \caption{Before and after steganography - Cosmic}
\end{figure}

\begin{figure}[!ht]
    \centering
    \begin{minipage}[b]{0.5\textwidth}
        \centering
        \includegraphics[width=0.85\textwidth]{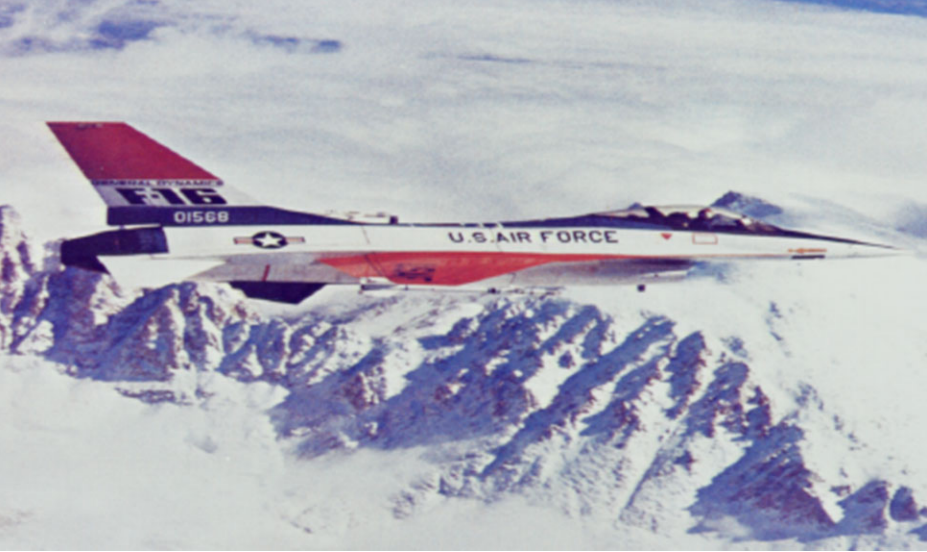}
        \caption{Before}
    \end{minipage}
    \begin{minipage}[b]{0.5\textwidth}
        \centering
        \includegraphics[width=0.85\textwidth]{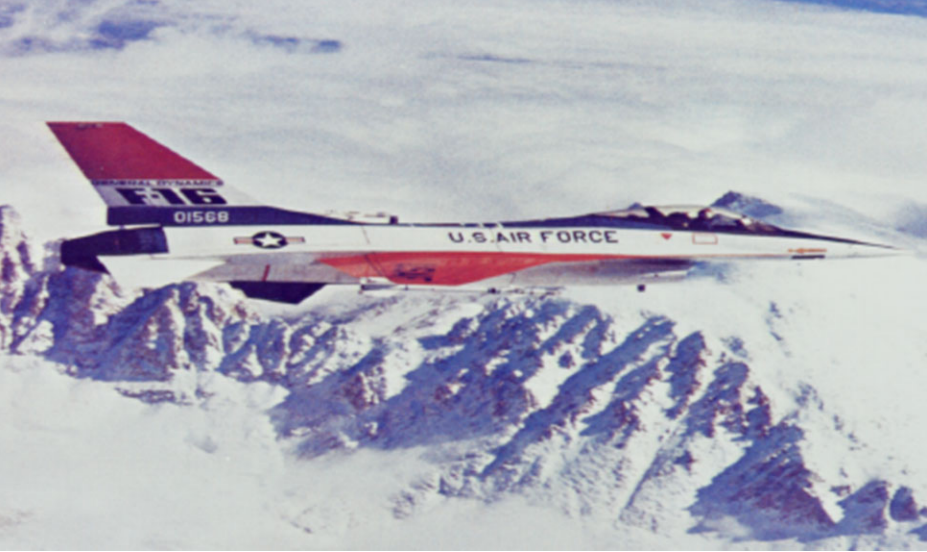}
        \caption{After}
    \end{minipage}
    \caption{Before and after steganography - F16}
\end{figure}

\begin{figure}[!ht]
    \centering
    \begin{minipage}[b]{0.5\textwidth}
        \centering
        \includegraphics[width=0.85\textwidth]{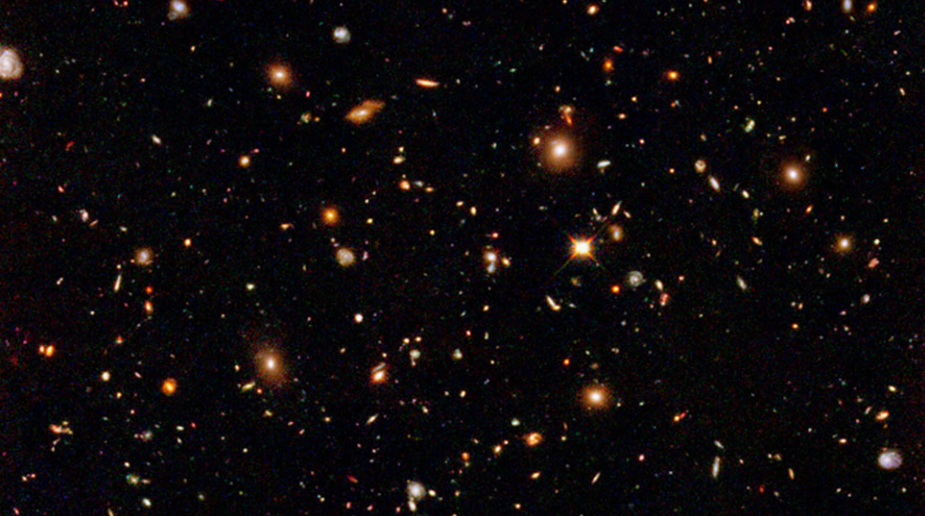}
        \caption{Before}
    \end{minipage}
    \begin{minipage}[b]{0.5\textwidth}
        \centering
        \includegraphics[width=0.85\textwidth]{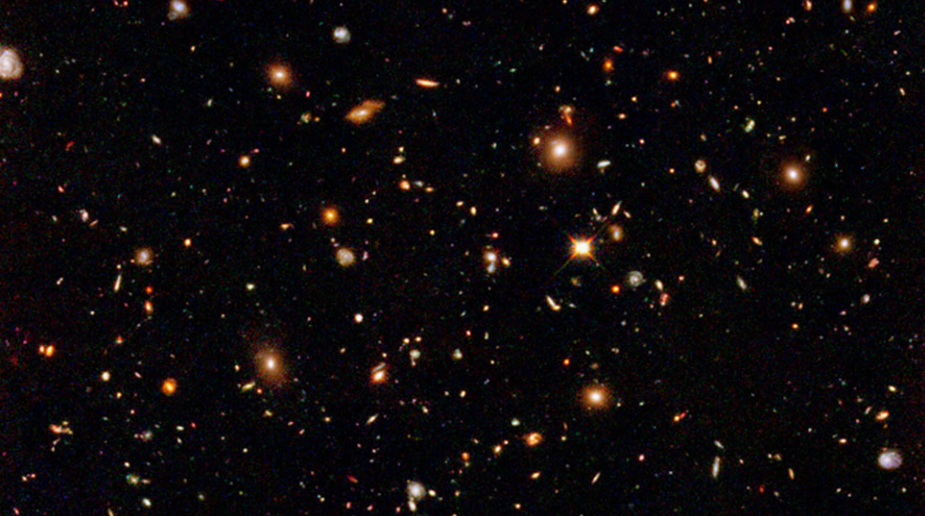}
        \caption{After}
    \end{minipage}
    \caption{Before and after steganography - Stars}
\end{figure}

\begin{figure}[!ht]
    \centering
    \begin{minipage}[b]{0.5\textwidth}
        \centering
        \includegraphics[width=0.85\textwidth]{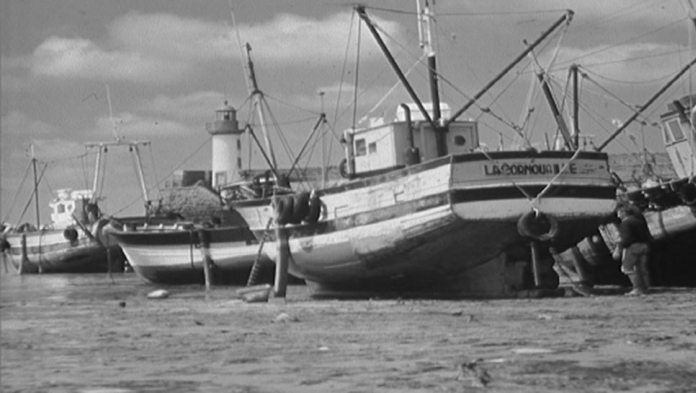}
        \caption{Before}
    \end{minipage}
    \begin{minipage}[b]{0.5\textwidth}
        \centering
        \includegraphics[width=0.85\textwidth]{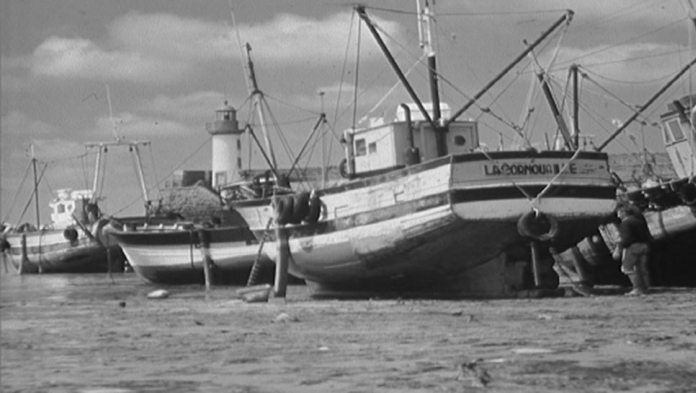}
        \caption{After}
    \end{minipage}
    \caption{Before and after steganography - Ship}
\end{figure}

\begin{figure*}
    \centering
    \includegraphics[width=\linewidth]{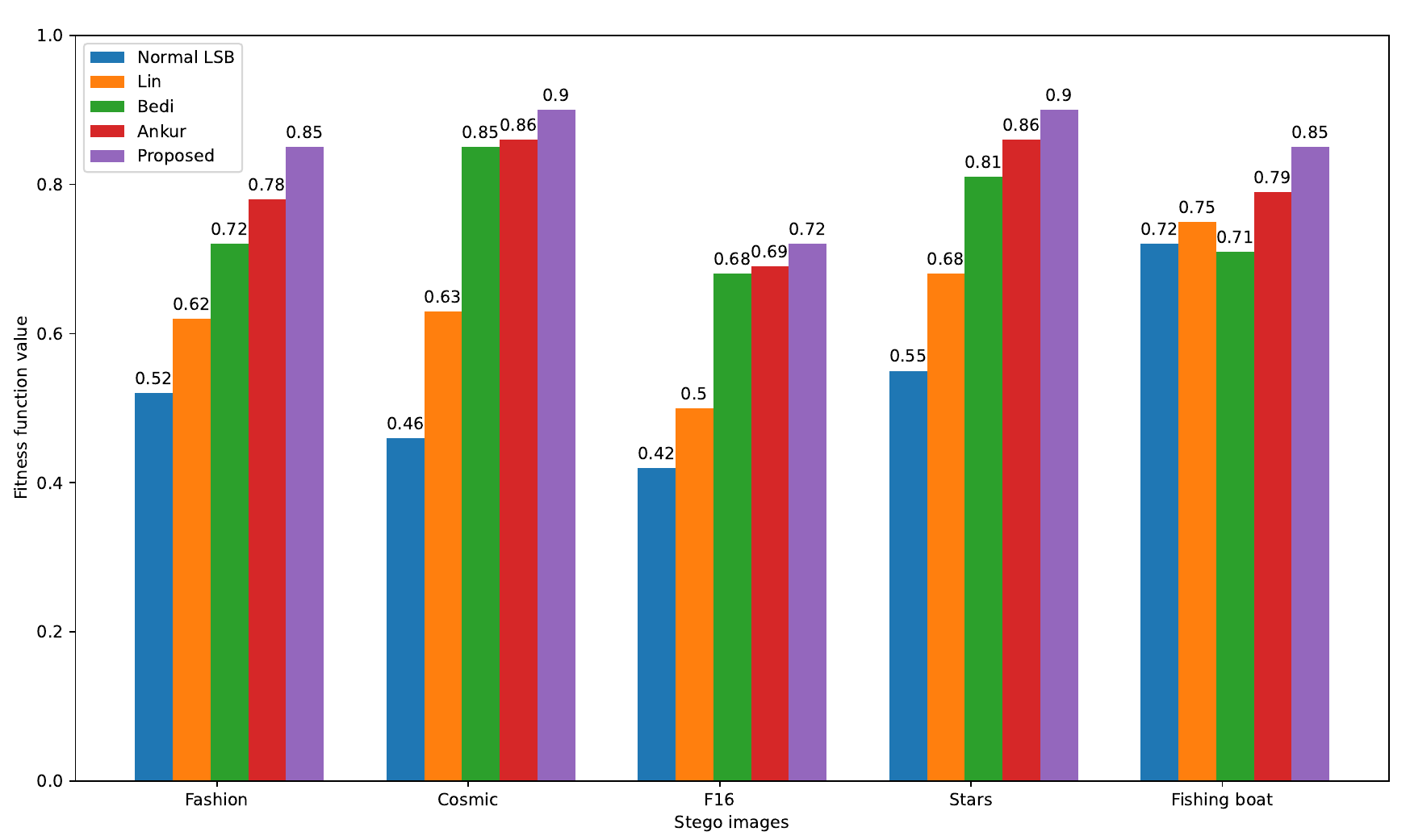}
    \caption{Comparing results of different SOTA steganographic methods}
\end{figure*}

\section{Experimental Results}\label{sec:experimental_results} 
Extensive experiments have been conducted to evaluate the proposed method's performance and the proposed method has shown state-of-the-art results in hiding an audio secret data in an image. We have compared our results with state-of-the-art steganographic methods. The results figure shows the comparison of Normal LSB, \cite{bedi2013using}, \cite{lin2009hiding}, \cite{gupta2018metaheuristic} and our proposed method. 

\clearpage
\section{Conclusion}\label{sec:conclusion}
In this work, we propose a unique method for concealing audio data inside otherwise unnoticeable image files, one that achieves outstanding results with minimal degradation of the original images. In order to obtain the best possible sequence of pixel locations to hide the secret audio data, the SFLA algorithm has played a crucial role. By mimicking frog leaping behavior, the SFLA algorithm search makes use of nature's evolutionary mechanisms to quickly arrive at the optimal solution. In keeping with the practice of most pixel replacement techniques, we have ensured that even after embedding audio in the host image, the overall size is not increased. When compared to other state-of-the-art approaches that aim to conceal binary audio data in an image, the proposed method excels in both the speed with which the optimal/sub-optimal solution is reached and the low degree of distortion introduced to the host image.

\section*{Acknowledgment}
We want to thank Dr. Nirman Kumar (nkumar8@memphis.edu) for insightful discussions.


\bibliographystyle{ieeetr}
\bibliography{bib_file}
\end{document}